# A peak in the critical current for quantum critical superconductors

Soon-Gil Jung[1], Soonbeom Seo[1], Sangyun Lee[1], Eric D. Bauer[2], Han-Oh Lee[3], and Tuson Park[1]*

1. Center for Quantum Materials and Superconductivity (CQMS), Department of Physics, Sungkyunkwan University, Suwon 16419, South Korea
2. Los Alamos National Laboratory, Los Alamos, NM 87545, USA
3. Center for Correlated Matter and Department of Physics, Zhejiang University, Hangzhou, Zhejiang 310058, China

## Abstract

Generally, studies of the critical current $I_c$ are necessary if superconductors are to be of practical use because $I_c$ sets the current limit below which there is a zero-resistance state. Here, we report a peak in the pressure dependence of the zero-field $I_c$, $I_c(0)$, at a hidden quantum critical point (QCP), where a continuous antiferromagnetic transition temperature is suppressed by pressure toward 0 K in CeRhIn$_5$ and 4.4% Sn-doped CeRhIn$_5$. The $I_c(0)$s of these Ce-based compounds under pressure exhibit a universal temperature dependence, underlining that the peak in zero-field $I_c(P)$ is determined predominantly by critical fluctuations associated with the hidden QCP. The dc conductivity $\sigma_{dc}$ is a minimum at the QCP, showing anti-correlation with $I_c(0)$. These discoveries demonstrate that a quantum critical point hidden inside the superconducting phase in strongly correlated materials can be exposed by the zero-field $I_c$, therefore providing a direct link between a QCP and unconventional superconductivity.

Unconventional superconductivity often is observed in close proximity to a magnetically-ordered phase, where the superconductivity (SC) transition temperature $T_c$ forms a dome against a non-thermal control parameter, such as the external pressure, chemical substitution, or magnetic field[1-6]. At an optimal value of the tuning parameter where $T_c$ is the highest, normal state properties do not follow predictions for Landau-Fermi liquids: the electrical resistivity ($\rho$) does not exhibit a $T^2$ dependence, and the electronic specific heat coefficient ($\gamma = C/T$) does not saturate but rather diverges with decreasing temperature[1, 2, 7]. These non-Fermi liquid (NFL) behaviours arise from incoherent critical fluctuations associated with a quantum critical point (QCP) hidden inside the SC dome of heavy-fermion compounds and some Fe-based superconductors such as $BaFe_2(As_{1-x}P_x)_2$ (refs. 1, 2, 4, 6, 8). Because the zero-temperature quantum phase transition typically is not accessible without destroying superconductivity, the role of critical magnetic fluctuations on properties of unconventional superconductors has yet to be explored in depth.

The critical current ($I_c$), which limits the current capacity of a zero-resistance state, characteristically is taken to depend on the strength of vortex pinning, which, in turn, is determined by the geometry and distribution of microstructural defects[9-11]. Because application of pressure should not lead to the creation of different or additional defects or to a substantial change in sample dimensions, $I_c$ in relation to $T_c$ should be at most weakly pressure dependent. A substantial variation in $I_c(P)$ or $I_c/T_c(P)$, then, logically, should be attributed to intrinsic changes in the superconducting state itself. For example, the zero-field critical current density $J_c$ (equal to $I_c/A$, where $A$ is the sample cross sectional area perpendicular to current) of the hole-doped high-$T_c$ cuprate superconductor $Y_{0.8}Ca_{0.2}Ba_2Cu_3O_y$ has a sharp peak that is centred on a critical hole-doping where the pseudogap boundary line projects to zero temperature and that is attributed in model calculations to changes in the superfluid density[12, 13]. These results indicate that $I_c$ measurements may provide an opportunity to explore the relationship between unconventional superconductivity and any QCP that is hidden beneath the SC dome.

Here, we report a peak in the zero-field critical current, $I_c(0)$, at a critical pressure $P_c$ in pure CeRhIn$_5$ (Rh115) and 4.4% Sn-doped CeRhIn$_5$ (SnRh115), where their respective antiferromagnetic boundary $T_N(P)$ extrapolates to $T = 0$ K inside a dome of pressure-induced superconductivity. The temperature dependence of $I_c(0)$s for pure Rh115 and SnRh115 under pressure is similar to that of superconducting CeCoIn$_5$, which is close to quantum criticality at ambient pressure. Normalized values of $I_c(T,P)$ follow a common universal curve for each material, suggesting an intrinsic, fundamental connection to quantum criticality. Supporting this conclusion, the magnetic field dependence of the flux-pinning force ($F_p = I_c \times \mu_0 H$), normalized to its maximum value, also forms a pressure-invariant universal curve for each compound. As will be discussed, these discoveries demonstrate that the pressure evolution of zero-field $I_c$ is determined mainly by quantum critical fluctuations and that the peak in $I_c$ is a direct link to the hidden QCP.

## Results

**Temperature-pressure phase diagrams.** Figures 1a and b present a contour plot of the zero-field $I_c(P,T)$ in the SC phase and the in-plane resistivity $\rho_{ab}(P,T)$ in the normal state for pure CeRhIn$_5$ (Rh115) and 4.4% Sn-doped CeRhSn$_{0.22}$In$_{4.78}$ (SnRh115) single crystals. The dependence on pressure of the in-plane resistivity and current-voltage curves upon which Fig. 1 is based is displayed in Supplementary Figs. 1 and 2, respectively. The quantum critical region veiled by the superconducting phase is fully exposed by the pressure dependence of zero-field

$I_c(T)$. A sharp peak in the value of $I_c(T)$ is clearly observed for pressures around the QCP at $P_c$, where a large enhancement in the resistivity is accompanied by strong quantum fluctuations[3, 4, 14]. In addition, $I_c(P)$ abruptly increases at pressures around $P_c^*$, the critical pressure where coexisting phases of magnetism and superconductivity evolve into a single SC state. In undoped Rh115, large differences between $T_c$s measured by heat capacity ($C$) and resistivity ($\rho$) at pressures below $P_c^*$ are ascribed to textured superconductivity originating from an incommensurate long-range magnetic order[15-18].

**Temperature dependences of the zero-field critical current.** The antiferromagnetic transition temperature ($T_N \sim 3.8$ K) in pure Rh115 is suppressed by Sn doping, which induces a shift of its extrapolated $T = 0$ K antiferromagnetic transition, and pressure-induced superconductivity emanates from the tuned QCP[4, 5] (see Supplementary Fig. 3). Figures 2a and b show the temperature dependence of the zero-field critical current, $I_c(0)$, for Rh115 and SnRh115 at several pressures, respectively. Here, $I_c$ is determined by using the voltage criterion of 0.1 $\mu$V (see Supplementary Fig. 2). Analysis of the flux-pinning force, $F_p = I_c \times \mu_0 H$, shows that the normalized flux-pinning force follows a power-law dependence on magnetic field, $f_p(h) \propto h^p(1-h)^q$, and is peaked around $h_{peak} \approx 0.6$, which is characteristic of type-II superconductors with weak pinning (see Supplementary Fig. 4). Here, the normalized pinning force is $f_p = F_p/F_{p,max}$ and the reduced field is $h = H/H_{irr}$, where $F_{p,max}$ is the maximum flux-pinning force and $H_{irr}$ is the irreversible field. The dependence on temperature of the critical current has been widely explained by $I_c(t)/I_c(0) = (1-t^2)^\alpha(1+t^2)^\beta$ for type-II superconductors, such as high-$T_c$ cuprates, Fe-based superconductors, and MgB$_2$ (refs. 19-22), where $t = T/T_c$ is the reduced temperature. When $T_c$ variations surrounding defects are important ($\delta T_c$-pinning), $\alpha = 7/6$ and $\beta = 5/6$, but $\alpha = 5/2$ and

$\beta$ = -1/2 for $\delta l$-pinning that arises from spatial variations in the charge-carrier mean free path ($l$) near a lattice defect[10, 19, 23]. These functional forms are shown by the dotted and dashed lines for $\delta T_c$- to $\delta l$-pinning in Fig. 2c, respectively. A crossover of the mechanism from $\delta T_c$- to $\delta l$-pinning has been often reported by introducing additional defects via chemical substitution or heavy ion irradiation, indicating that $\delta T_c$-pinning is preferred in clean crystals[19-21].

The values of $I_c(t)$ for SnRh115 at pressures around $P_c$ are fitted together with those for Rh115 and CeCoIn$_5$ in Fig. 2c. The temperature dependence of $I_c(0)$ for CeCoIn$_5$ at ambient pressure is measured to compare it with that of CeRhIn$_5$ because Rh115 is believed to have a superconducting pairing mechanism similar to that in CeCoIn$_5$. The values of $I_c(0)$ for all samples can be expressed well by one curve with the relation $I_c(t) \propto (1-t^2)^{5/6}(1+t^2)^{2/3}$, which is distinct from that for $I_c(t)$ controlled by either $\delta T_c$- or $\delta l$-pinning. This universal curve underscores that the origin of the zero-field $I_c$ is the same for each compound and that it does not change under pressure for these Ce-based quantum critical materials. The fact that external pressure does not create new defects inside the crystals suggests that the pressure evolution of $I_c$ should be related to the pressure dependence of the superconducting coupling strength.

**Discussion**

Figure 3a presents the pressure dependences of $I_c(0)$ and $T_{c,0}$ for SnRh115, which are similar to each other. However, their relative fractional variations in $I_c$ and $T_c$, $\gamma_I \equiv I_{c,0}(P)/I_{c,0}(P_c) \times 100$ and $\gamma_T \equiv T_{c,0}(P)/T_{c,0}(P_c) \times 100$, where $I_{c,0}(P_c)$ is $I_c$ extrapolated to zero temperature at $P_c$ and $T_{c,0}(P_c)$ is the superconducting transition temperature at $P_c$, are much different, as shown in Fig. 3b: at 1.0 GPa, the critical current is 45% of the maximum value, and $T_{c,0}$ is 67% of its maximal value. The stronger pressure dependence of $I_c$ relative to that of $T_{c,0}$ is clearly visible in ratio $I_{c,0}/T_{c,0}$ for

SnRh115, as presented in Figs. 3c and d. An abrupt enhancement in $I_{c,0}/T_{c,0}$ is observed at $P_c^*$, and the peak in the pressure dependence of $I_{c,0}/T_{c,0}$ is achieved at $P_c$. The dc conductivity (= $\sigma_{dc}$) at $T_c$ onset is shown as a function of pressure in the right ordinate of Fig. 3c, where a minimum value appears near $P_c$ (see Supplementary Fig. 5). The anti-correlation between $I_{c,0}/T_{c,0}$ and $\sigma_{dc}$ in these Ce-based quantum critical compounds may be related with the presence of the hidden QCP at $P_c$ because the associated critical quantum fluctuations not only act as the SC pairing glue but also strongly enhance incoherent electron scattering, thus leading to a minimum in $\sigma_{dc}$ at $P_c$[24, 25]. Homes' scaling relation[26-28] states that the superfluid density $n_s$ is proportional to $\sigma_{dc}T_c$ in many correlated superconductors and, consequently, that the ratio $n_s/T_c$ should be proportional to $\sigma_{dc}$. The fact that $\sigma_{dc}$ is the minimum at $P_c$ where $I_{c,0}/T_{c,0}$ is the maximum in these Ce-based compounds suggests a violation of Homes' scaling if the strength of the condensate $n_s$ is proportional to the critical current $I_{c,0}$. Pressure-dependent optical conductivity and/or penetration depth experiments that directly measure $n_s$ will be important to provide a stringent test for the validity of Homes' law in quantum critical superconductors.

Our study demonstrates that the critical current, a fundamental superconducting parameter, is a powerful tool for investigating the presence of a hidden quantum critical point inside the superconducting dome without destroying the superconducting phase. The dependence on temperature of the zero-field $I_c$ for both pure Rh115 and Sn-doped Rh115 exhibits the same functional form under pressure, underscoring that the peak at $P_c$ in the pressure dependence of $I_c$ arises from an enhanced fluctuations around the hidden quantum critical point. Even though these results are specific to the Ce115 heavy fermion materials, the prediction of similar results for the hole-doping dependence of the critical current density $J_c(x)$ in high-$T_c$ cuprates[29] suggests a universal behaviour of $J_c$ among unconventional superconductors. These discoveries should stimulate more theoretical and experimental effort to understand the intimate link between quantum criticality and the origin of unconventional superconductivity in various families of correlated electronic systems.

## Methods

**Measurement outline.** CeRhIn$_5$, Sn-doped CeRhIn$_5$, and CeCoIn$_5$ single crystals were synthesized by the indium (In) self-flux method[30-32]. Pressure was generated in a hybrid clamp-type pressure cell with Daphne 7373 as the pressure-transmitting medium, and the pressure was determined by monitoring the shift in the value of $T_c$ for lead (Pb). Measurements of current – voltage ($I – V$) characteristics under pressure were performed in a Heliox VL system (Oxford Instruments) with a vector magnet ($y$ = 5 T and $z$ = 9 T, American Magnetics Inc.) and in a Physical Property Measurement System (PPMS 9 T, Quantum Design), where the current was provided by a Keithley 6221 unit and the voltage was measured with a Keithley 2182A nanovoltmeter.

**Measurement details.** Measurements of $I – V$ characteristics were performed in a pulsed mode to minimize Joule heating developed at Ohmic contacts to the samples and copper (Cu) wires between the pressure cell and the connector. The duration of the pulsed current was 10 - 11 ms, and the repetition rate was one pulse every 2 s, which was sufficient to eliminate Joule heat in the samples[33, 34]. A standard four-probe method was used to determine $I – V$, and good Ohmic contact to samples was achieved by using silver epoxy. The critical current was based on a $10^{-7}$ $V$ criterion[35], which was averaged over three measurements. The dimensions of the measured crystals were 920 × 330 × 20 $\mu$m$^3$, 650 × 200 × 22 $\mu$m$^3$, and 1100 × 200 × 47 $\mu$m$^3$ for CeRhIn$_5$ (Rh115), CeRhSn$_{0.22}$In$_{4.78}$ (SnRh115), and CeCoIn$_5$, respectively. The magnetic-field dependence of the critical current was measured at several pressures and the flux-pinning force ($F_p$) was estimated from the relation $F_p = I_c \times \mu_0 H$ (refs. 36-38).

**Data availability.** The data sets generated and/or analyzed in this study are available from the corresponding author on reasonable request.

**References:**

1. Park, T. et al. Hidden magnetism and quantum criticality in the heavy fermion superconductor CeRhIn$_5$. *Nature* **440**, 65-68 (2006).

2. Knebel, G., Aoki, D., Braithwaite, D., Salce, B. & Flouquet, J. Coexistence of antiferromagnetism and superconductivity in CeRhIn$_5$ under high pressure and magnetic field. *Phys. Rev. B* **74**, 020501(R) (2006).

3. Park, T. et al. Isotropic quantum scattering and unconventional superconductivity. *Nature* **456**, 366-368 (2008).

4. Seo, S. et al. Controlling superconductivity by tunable quantum critical points. *Nat. Commun.* **6**, 6433 (2015).

5. Ferreira, L. M. et al. Tuning the pressure-induced superconducting phase in doped CeRhIn$_5$. *Phys. Rev. Lett.* **101**, 017005 (2008).

6. Analytis, J. G. et al. Transport near a quantum critical point in BaFe$_2$(As$_{1-x}$P$_x$)$_2$, *Nat. Phys.* **10**, 194-197 (2014).

7. Gegenwart, P., Si, Q. & Steglich F. Quantum criticality in heavy-fermion metals. *Nat. Phys.* **145**, 186-197 (2008).

8. Hashimoto, K. et al. A sharp peak of the zero-temperature penetration depth at optimal composition in BaFe$_2$(As$_{1-x}$P$_x$)$_2$, Science **336**, 1554-1557 (2012).

9. Fang, L. et al. Huge critical current density and tailored superconducting anisotropy in SmFeAsO$_{0.8}$F$_{0.15}$ by low-density columnar-defect incorporation, *Nat. Commun.* **4**, 2655 (2013).

10. Kunchur, M. N., Lee, S.-I. & Kang, W. N. Pair-breaking critical current density of magnesium diboride. *Phys. Rev. B* **68**, 064516 (2003).

11. Dew-Hughes, D. The critical current of superconductors: an historical review. *Low Temp. Phys.* **27**, 713-722 (2001).

12. Talantsev, E. F. & Tallon, J. L. Universal self-field critical current for thin-film superconductors. *Nat. Commun.* **6**, 7820 (2015).

13. Talantsev, E. F., Crump, W. P. & Tallon, J. L. Thermodynamic parameters of single- or multi-band superconductors derived from self-field critical currents. *Ann. Phys.* (Berlin) 1700197 (2017).

14. Knebel, G., Aoki, D., Brison, J.-P. & Flouquet, J. The quantum critical point in CeRhIn$_5$: A resistivity study. *J. Phys. Soc. Jpn.* **77**, 114704 (2008).

15. Park, T. et al. Textured superconducting phase in the heavy fermion CeRhIn$_5$. *Phys. Rev. Lett*. **108**, 077003 (2012).

16. Park, T. & Thompson, J. D. Magnetism and superconductivity in strongly correlated CeRhIn$_5$. *New. J. Phys.* **11**, 055062 (2009).

17. Llobert, A. et al. Magnetic structure of CeRhIn$_5$ as a function of pressure and temperature, *Phys. Rev. B* **69**, 024403 (2004).

18. Yashima, M. et al. Strong coupling between antiferromagnetic and superconducting order parameters of CeRhIn$_5$ studied by $^{115}$In nuclear quadrupole resonance spectroscopy. *Phys. Rev. B* **79**, 214528 (2009).

19. Griessen, R. et al. Evidence for mean free path fluctuation induced pinning in YBa$_2$Cu$_3$O$_7$ and YBa$_2$Cu$_4$O$_8$ films. *Phys. Rev. Lett*. **72**, 1910-1913 (1994).

20. Wen, H. H., Zhao, Z. X., Xiao, Y. G., Yin, B. & Li, J. W. Evidence for flux pinning induced by spatial fluctuation of transition temperatures in single domain (Y$_{1-x}$Pr$_x$)Ba$_2$Cu$_3$O$_{7-\delta}$ samples. *Physica* C **251**, 371-378 (1995).

21. Xiang, F. X. et al. Evidence for transformation from $\delta T_c$ to $\delta l$ pinning in MgB$_2$ by graphene oxide doping with improved low and high field $J_c$ and pinning potential. *Appl. Phys. Lett*. **102**, 152604 (2013).

22. Ghorbani, S. R., Wang, X. L., Shahbazi, M., Dou. S. X. & Lin, C. T. Fluctuation of mean free path and transition temperature induced vortex pinning in (Ba,K)Fe$_2$As$_2$ superconductors. *Appl. Phys. Lett*. **100**, 212601 (2012).

23. Blatter, G., Feigel'man, M. V., Geshkenbein, V. B., Larkin, A. I. & Vinokur, V. M. Vortices in high-temperature superconductors. *Rev. Mod. Phys*. **66**, 1125 (1994).

24. Howald, L., Knebel, G., Aoki, D., Lapertot, G. & Brison, J.-P. The upper critical field of CeCoIn$_5$. *New J. Phys*. **13**, 113039 (2011).

25. Miyake, K. & Narikiyo, O. Enhanced impurity scattering due to quantum critical fluctuations: perturbational approach. *J. Phys. Soc. Jpn*. **71**, 867-871 (2002).

26. Homes, C. C. et al. A universal scaling relation in high-temperature superconductors. *Nature* **430**, 539-541 (2004).

27. Dordevic, S. V., Basov, D. N. & Homes, C. C. Do organic and other exotic superconductors fail universal scaling relations? *Sci. Rep*. **3**, 1713 (2013).

28. Uemura, Y. J. et. al. Universal correlations between $T_c$ and $n_s/m^*$ (carrier density over effective mass) in high-$T_c$ cuprate superconductors. *Phys. Rev. Lett*. **62**, 2317-2320 (1989).

29. Tallon, J. L. et al. Critical doping in overdoped high-$T_c$ superconductors: a quantum critical point? *Phys. Stat. Sol*. (b) **251**, 531-540 (1999).

30. Hegger, H. et al. Pressure-induced superconductivity in quasi-2D $CeRhIn_5$. *Phys. Rev. Lett*. **84**, 4986-4989 (2000).

31. Bauer, E. D. et al. Antiferromagnetic quantum critical point in $CeRhIn_{5-x}Sn_x$. *Physica* B **378-380**, 142-143 (2006).

32. Petrovic, C. et al. Heavy-fermion superconductivity in $CeCoIn_5$ at 2.3 K. *J. Phys.: Condens. Matter* **13**, L337-L342 (2001).

33. Kunchur, M. N. Current-induced pair breaking in magnesium diboride. *J. Phys.: Condens. Matter* **16**, R1183-R1204 (2004).

34. Liang, M., Kunchur, M. N., Fruchter, L. & Li, Z. Z. Depairing current density of infinite-layer $Sr_{1-x}La_xCuO_2$ superconducting films. *Physica* C **492**, 178-180 (2013).

35. Dobrovolskiy, O. V., Begun, E., Huth, M. & Shklovskij, V. A. Electrical transport and pinning properties of Nb thin films patterned with focused ion beam-milled washboard nanostructures. *New J. Phys*. **14**, 113027 (2012).

36. Fietz, W. A. & Webb, W. W. Hysteresis in superconducting alloys-temperature and field dependence of dislocation pinning in niobium alloys. *Phys. Rev*. **178**, 657-667 (1969).

37. Dew-Hughes, D. Flux pinning mechanisms in type-II superconductors. *Philos. Mag.* **30**, 293-305 (1974).

38. Kramer, E. J. Scaling laws for flux pinning in hard superconductors. *J. Appl. Phys*. **44**, 1360-1370 (1973).

**Acknowledgements:** We thank J. D. Thompson, Y. Bang, Z. Fisk, I. Vekhter, Y. Yang, J. H. Yun, and S. Oh for helpful discussions. This work was supported by a National Research Foundation (NRF) of Korea grant funded by the Korean Ministry of Science, ICT and Planning (No. 2012R1A3A2048816). S.-G. Jung was supported by the Basic Science Research Program through the National Research Foundation of Korea (NRF) funded by the Ministry of Education (NRF-2015R1D1A1A01060382). Work at Los Alamos National Laboratory was performed under the auspices of the U.S. Department of Energy, Office of Basic Energy Sciences, Division of Materials Sciences and Engineering.

**Author contributions:** S.-G.J. conceived the work. S.-G.J., S.S., and S.L. performed the $I-V$ measurements at various pressures. E.D.B. and H.-O.L. synthesized the $CeRhIn_5$, Sn-doped $CeRhIn_5$, and $CeCoIn_5$ single crystals. S.-G.J. analyzed the data and discussed the results with all authors. The manuscript was written by S.-G.J. and T.P. with inputs from all authors.

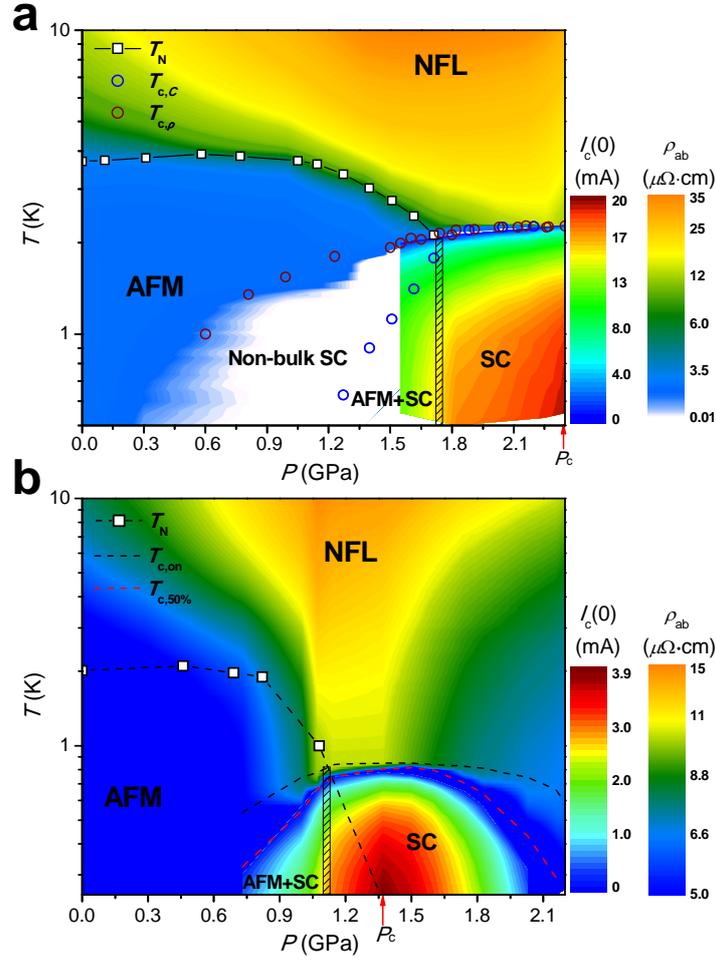

**Figure 1. Temperature-pressure phase diagrams of CeRhIn$_5$ and CeRhSn$_{0.22}$In$_{4.78}$ single crystals.** In the superconducting state below $T_c(P)$, false colours denote the magnitude of the zero-field critical current $I_c(P, T)$. At temperatures above $T_c(P)$, false colours reflect the magnitude of the in-plane resistivity $\rho_{ab}(P, T)$. **a,** CeRhIn$_5$ (Rh115) and **b,** CeRhSn$_{0.22}$In$_{4.78}$ (SnRh115). For both materials, $\rho_{ab}(P, T)$ is enhanced around the quantum critical point $P_c$ due to pronounced incoherent inelastic scattering. Similarly, the zero-field $I_c(P,T)$ is largest at $P_c$ where the QCP is expected, as indicated by the arrow. In both **a**, and **b**, the vertical hashed rectangle is at $P_c^*$, the pressure that separates a phase of coexisting superconductivity and magnetism from a purely superconducting phase for $P > P_c^*$. Open squares in both **a**, and **b** represent the antiferromagnetic transition temperature ($T_N$). Superconducting transition temperature ($T_c$) of Rh115 is evaluated from specific heat ($T_{c,C}$) and resistivity ($T_{c,\rho}$) measurements, and $T_c$ of SnRh115 is determined as $T_c$ onset ($T_{c,on}$) and 50% ($T_{c,50\%}$) of the normal-state resistivity value at $T_{c,on}$. AFM, SC, and NFL stand for antiferromagnetic, superconducting, and non-Fermi liquid regions, respectively.

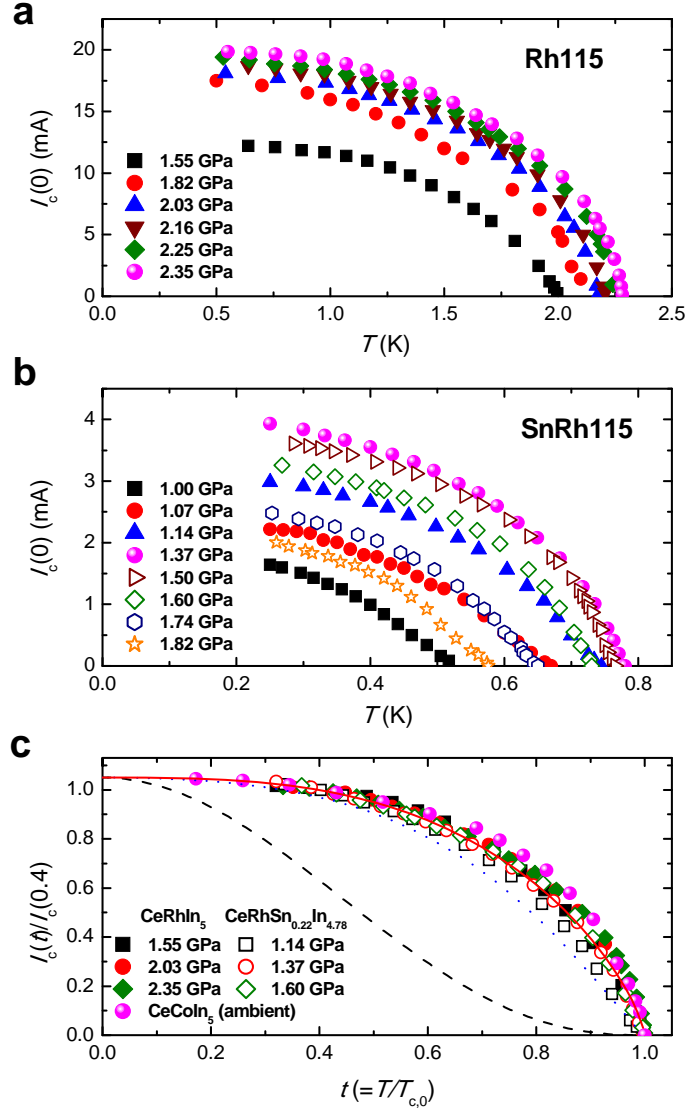

**Figure 2. Temperature dependences of the zero-field critical current for Ce-based heavy-fermion materials under pressure**. **a,** Temperature dependence of the zero-field critical current, $I_c(0)$, for CeRhIn$_5$ at various pressures. **b,** Zero-field $I_c$ for CeRhSn$_{0.22}$In$_{4.78}$ at various pressures. **c,** Reduced temperature ($t = T/T_{c,0}$) dependence of $I_c$, $I_c(t)$, normalized by its value at $t = 0.4$ for Rh115 and SnRh115 at representative pressures and for CeCoIn$_5$ at ambient pressure. The normalized values of $I_c(P,t)$ for all crystals can be described by a single curve, $I_c(t) \propto (1-t^2)^{5/6}(1+t^2)^{2/3}$, indicating universal behaviour of $I_c(t)$ with respect to pressure in the Ce$M$In$_5$ ($M$ = Co, Rh) materials. Dotted and dashed curves are for $\delta T_c$- and $\delta l$-pinning, respectively, as discussed in the text.

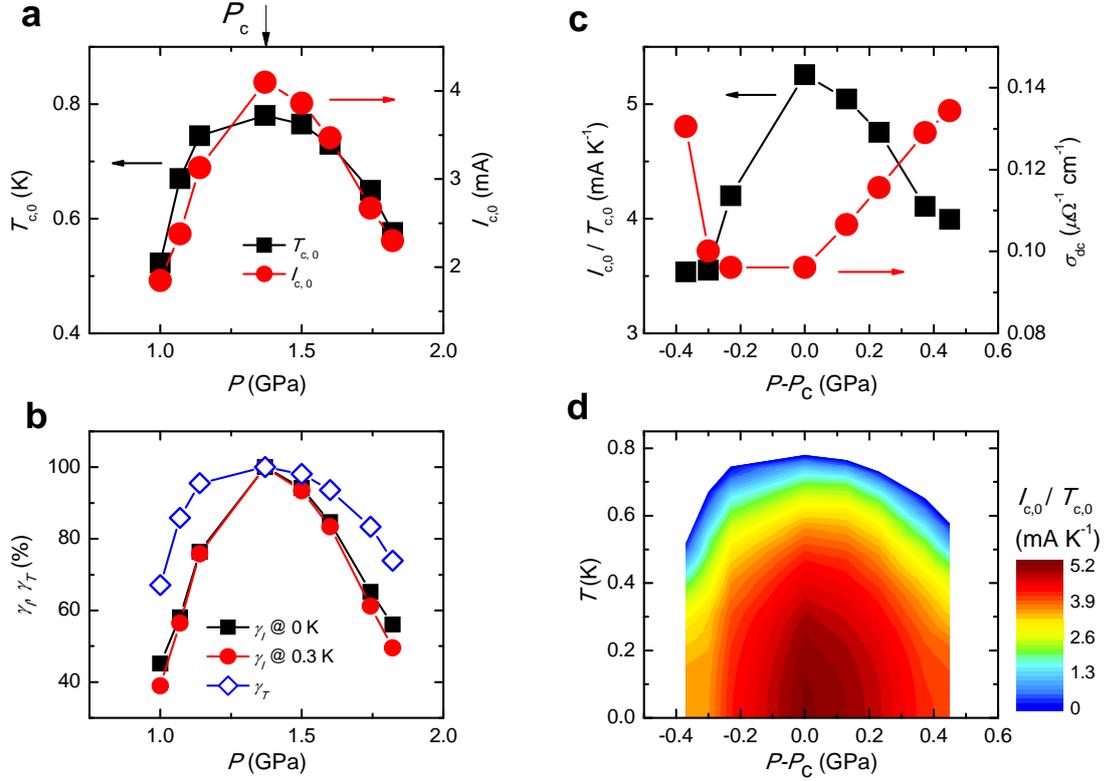

**Figure 3. Pressure evolution of the zero-field critical current in Sn-doped CeRhIn$_5$. a,** Pressure dependences of $T_{c,0}$ and $I_{c,0}$ for SnRh115, where $I_{c,0}$ is the value of $I_c$ obtained from an extrapolation of data in Fig. 2b to zero Kelvin. **b,** Fractional variations in $I_{c,0}$ and $T_{c,0}$ for SnRh115 under pressure. The fractions are defined as $\gamma_I \equiv I_c(T,P)/I_{c,0}(P_c) \times 100$ and $\gamma_T \equiv T_{c,0}(P)/T_{c,0}(P_c) \times 100$, where $I_{c,0}(P_c)$ is $I_c$ extrapolated to zero temperature at $P_c$ and $T_{c,0}(P_c)$ is the superconducting transition temperature at $P_c$. Values of $\gamma_I$ are plotted as a function of pressure for measured or estimated $I_c(T,P)$ at 0 K (squares) and 0.3 K (circles). **c,** The ratio between the critical current and SC transition temperature, $I_{c,0}/T_{c,0}$, plotted together with the dc conductivity at $T_c$ onset, $\sigma_{dc}$, as a function of the pressure difference $P$-$P_c$, where $P_c$ = 1.35 GPa is the QCP. **d,** A contour plot of $I_{c,0}/T_{c,0}$ displayed in the temperature ($T$) and pressure ($P$-$P_c$) plane. The ratio $I_{c,0}/T_{c,0}$ forms a dome centred around the quantum critical point $P_c$ and its values decrease with distance from $P_c$.

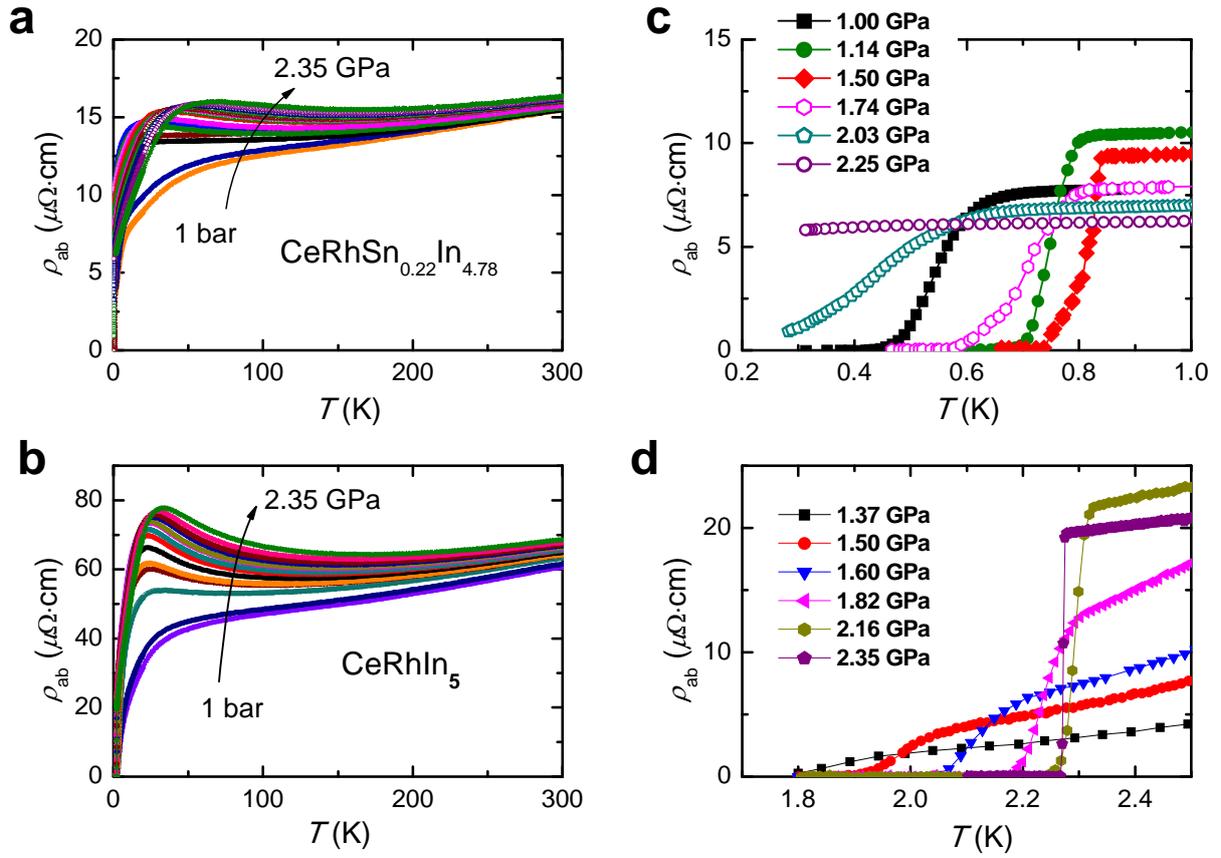

**Supplementary Figure 1: Temperature dependences of the in-plane resistivity for CeRhSn$_{0.22}$In$_{4.78}$ and CeRhIn$_5$ under pressure. a,** In-plane resistivity ($\rho_{ab}$) of the Sn-doped Rh115 (SnRh115) is plotted as a function of temperature for pressures, $\rho_{ab}(P,T)$, of 0, 0.14, 0.97, 1.0, 1.07, 1.14, 1.37, 1.5, 1.6, 1.74, 1.82, 2.03, 2.16, 2.25, and 2.35 GPa along the arrow direction. **b,** $\rho_{ab}(P,T)$ of undoped Rh115 is displayed against temperature for pressures of 0, 0.14, 0.97, 1.07, 1.14, 1.37, 1.5, 1.6, 1.74, 1.82, 2.03, 2.16, 2.25, and 2.35 GPa along the arrow direction. A magnified view of $\rho_{ab}(P,T)$ near the SC transition temperatures is shown at selected pressures for SnRh115 and Rh115 in **c** and **d**, respectively.

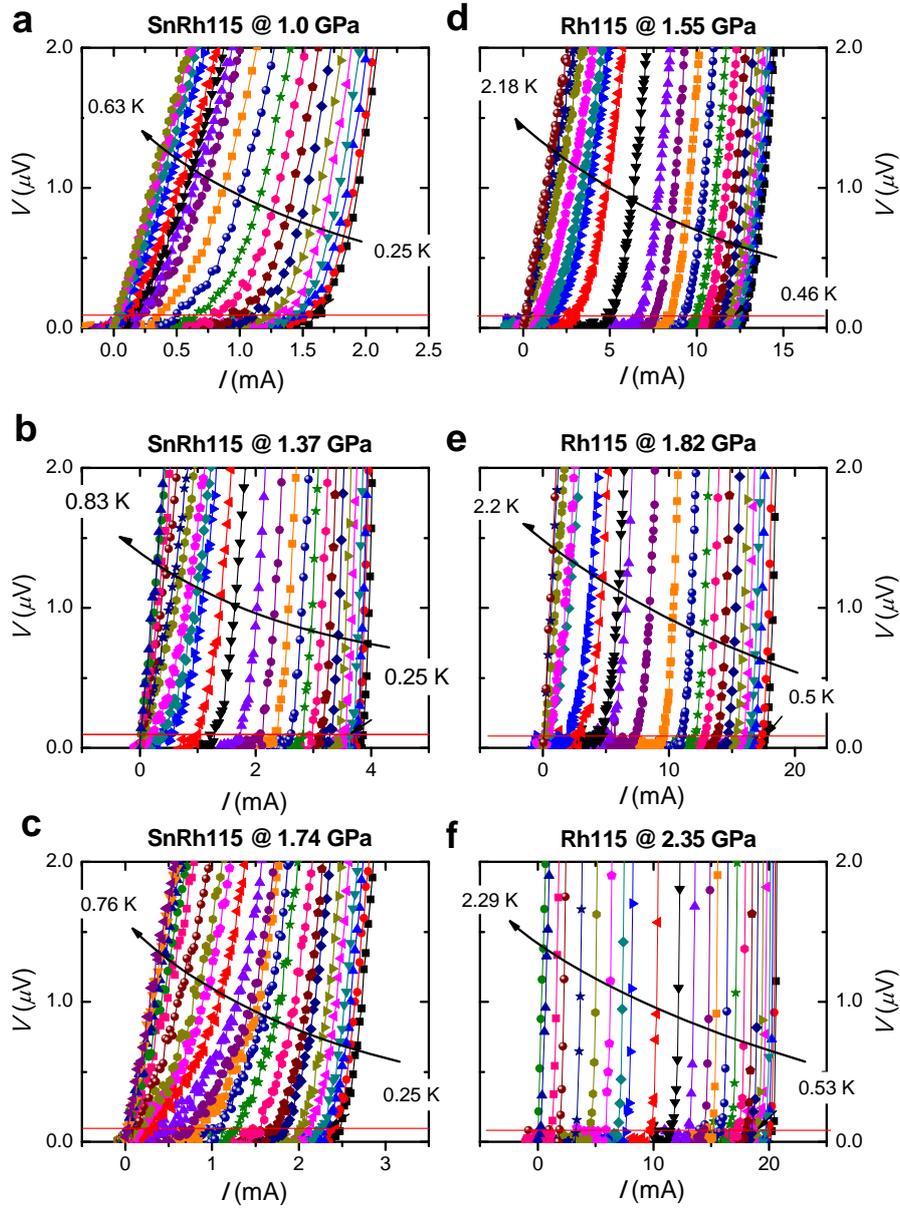

**Supplementary Figure 2: Current – voltage curves at zero field for CeRhSn$_{0.22}$In$_{4.78}$ and CeRhIn$_5$ under various pressures. a-c,** Current ($I$) – voltage ($V$) characteristic curves for the SnRh115 single crystal at pressures of 1.0, 1.37, and 1.74 GPa. **d-f,** $I – V$ characteristic curves for the Rh115 single crystal at pressures of 1.55, 1.82, and 2.35 GPa. The critical current ($I_c$) is determined using the $V = 0.1$ $\mu V$ criterion that is indicated as a horizontal red line. Arrows indicate the temperature variation over which $I – V$ curves were measured.

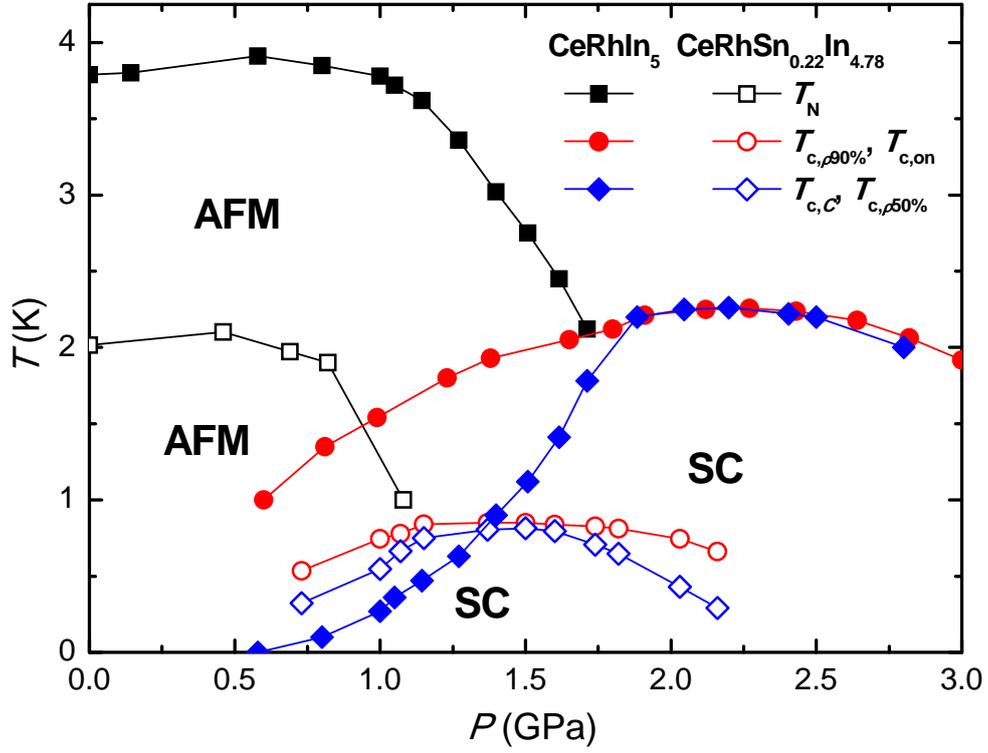

**Supplementary Figure 3: Temperature-pressure phase diagrams for CeRhIn$_5$ and CeRhSn$_{0.22}$In$_{4.78}$ single crystals.** Solid and open squares represent the antiferromagnetic (AFM) transition temperature ($T_N$) for Rh115 and SnRh115, respectively. Solid circles and solid diamonds describe the superconducting transition temperatures ($T_c$'s) evaluated from resistivity ($T_{c,\rho 90\%}$) and specific heat ($T_{c,C}$) measurements on Rh115, respectively. Data at pressures above 2.35 GPa are adapted from Ref. [1]. Open circles ($T_{c,on}$) and open diamonds ($T_{c,\rho 50\%}$) are $T_c$ of SnRh115 that were determined as $T_c$ onset and 50% of the normal-state resistivity value at $T_{c,on}$, respectively.

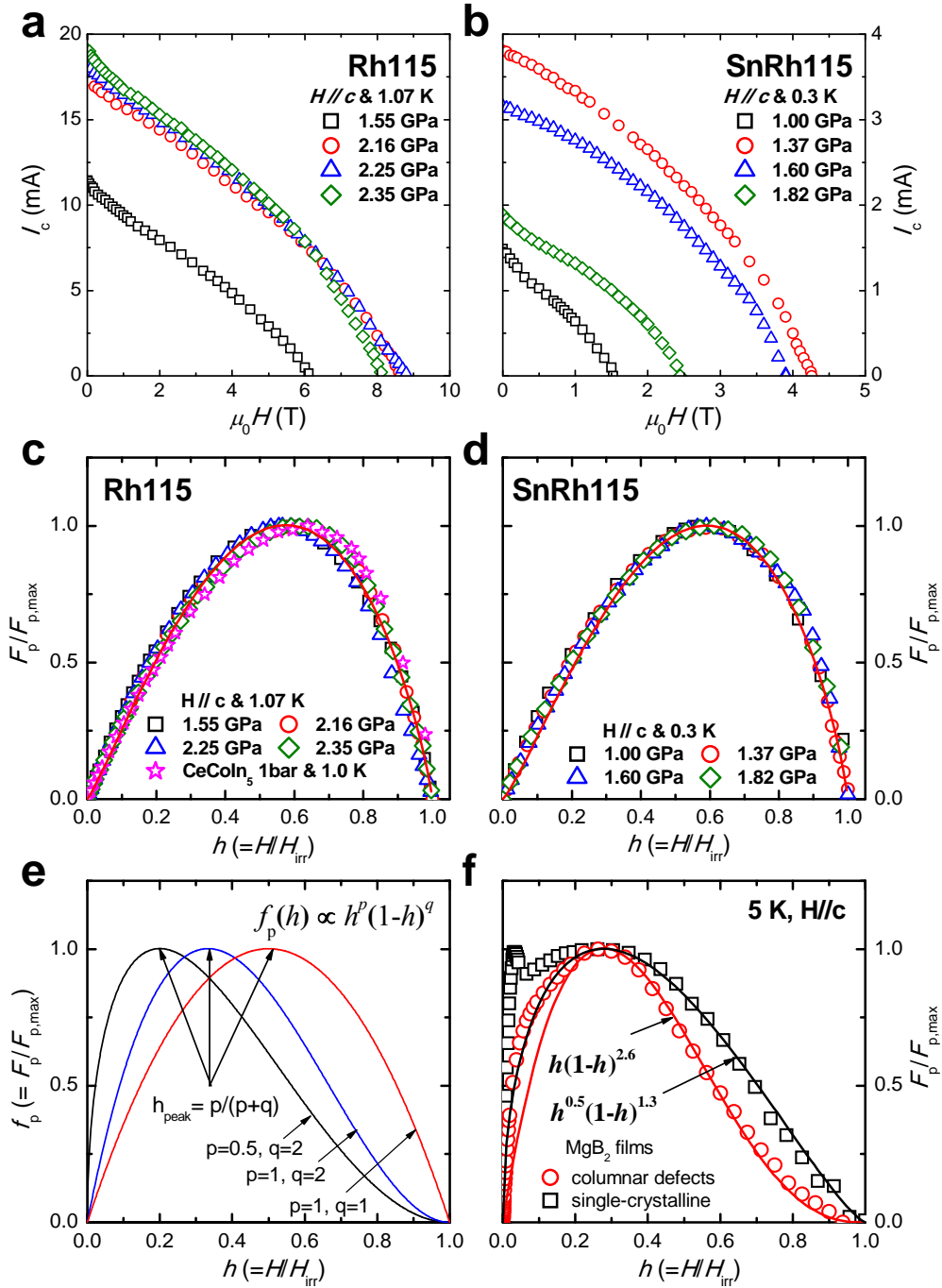

**Supplementary Figure 4: Magnetic field dependences of critical current and normalized flux-pinning force for CeRhIn$_5$ and CeRhSn$_{0.22}$In$_{4.78}$. a, b** show the critical current ($I_c$) as a function of magnetic fields for CeRhIn$_5$ (Rh115) and CeRhSn$_{0.22}$In$_{4.78}$ (SnRh115) under pressure, respectively. The flux pinning force ($F_p$) can be obtained from $I_c(H)$ using the relation $F_p = I_c \times \mu_0 H$. **c,** Reduced field ($h = H/H_{irr}$) dependence of the normalized flux-pinning force ($f_p = F_p/F_{p,max}$)

for Rh115 and CeCoIn$_5$, where $H_{irr}$ is the irreversible field and $F_{p,max}$ is the maximum flux-pinning force. All $f_p(h)$ sets of data for Rh115 are well scaled by one curve irrespective of the pressure, and the $f_p(h)$ curve for CeCoIn$_5$ at ambient pressure is similar to Rh115 under pressure. **d,** $f_p(h)$ curves for SnRh115 also are well expressed by one curve regardless of the pressure. The similar scaling relations for $f_p(h)$ for all samples indicate that the main flux-pinning sources in the samples are the same [2,3]. **e**, Several commonly used scaling curves $f_p(h)$ are plotted against reduced field $h$ [2,3]. The field dependence of the normalized flux-pinning force, $f_p(h)$, generally can be expressed by the relation $f_p(h) \propto h^p(1-h)^q$, and at low fields, the maximum flux-pinning force usually occurs at $h_{peak} = p/(p+q)$ in type-II superconductors with strong pinning. Phenomenologically, the fitting parameter $p$ is close to 2 for strong pinning and $p \approx 1$ for weak pinning. Our Ce-based 115 compounds, which are very clean materials, show a scaling behaviour $f_p(h)$ with $h_{peak} \approx 0.6$, which is plotted as a red line in **a** and **b**, where the best results were obtained with $p = 1.15$ and $q = 0.85$ for Rh115 and $p = 1.2$ and $q = 0.83$ for SnRh115. **f**, For comparison, $f_p(h)$ is shown as a function of reduced field for single-crystalline MgB$_2$ thin films and MgB$_2$ films with a strong flux-pinning source of a columnar grain boundary.

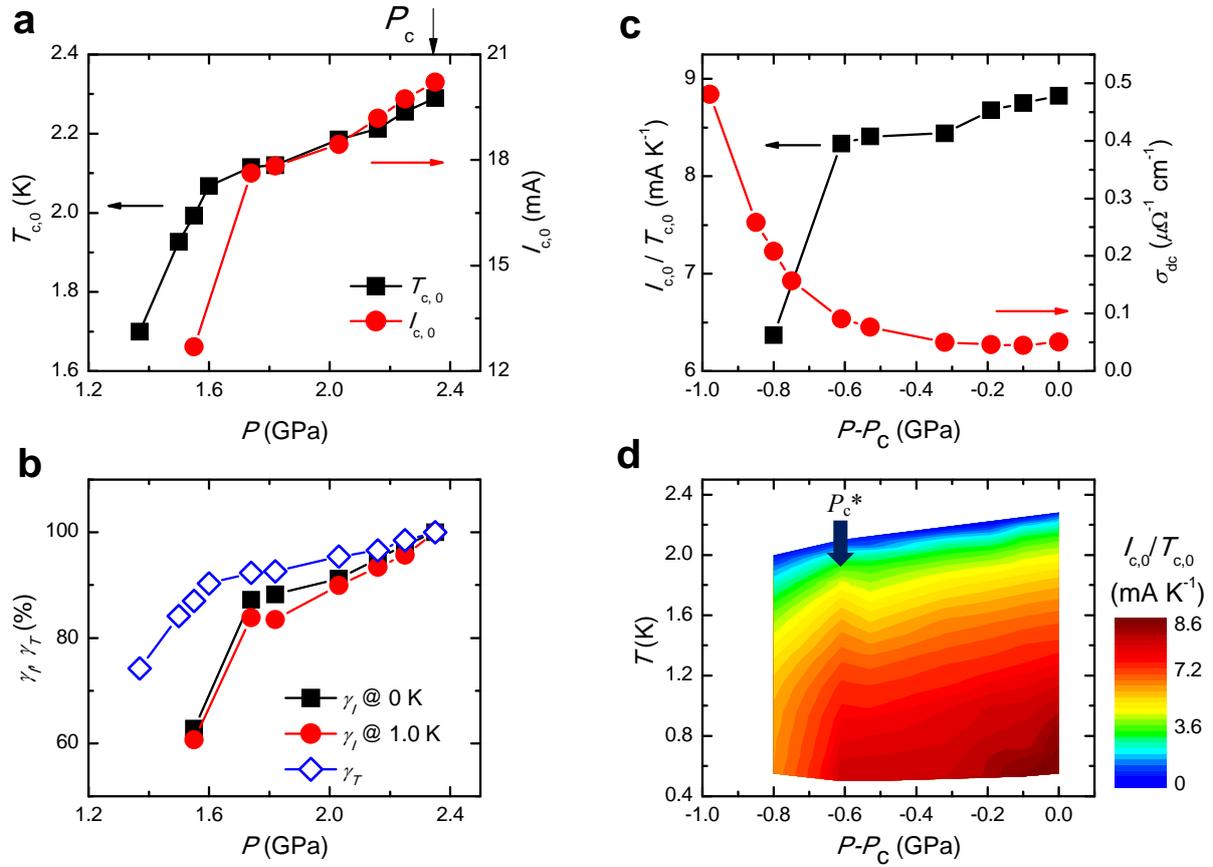

**Supplementary Figure 5: Pressure evolution of the zero-field critical current in CeRhIn$_5$. a,** Pressure dependences of $T_{c,0}$ and $I_{c,0}$ for Rh115, where $I_{c,0}$ is the value of $I_c$ obtained from an extrapolation of data in Fig. 2a to zero Kelvin. **b,** Fractional variations in $I_{c,0}$ and $T_{c,0}$ for Rh115 under pressure. The fractions are defined as $\gamma_I \equiv I_c(T,P)/I_{c,0}(P_c) \times 100$ and $\gamma_T \equiv T_{c,0}(P)/T_{c,0}(P_c) \times 100$, where $I_{c,0}(P_c)$ is $I_c$ extrapolated to zero temperature at $P_c$ and $T_{c,0}(P_c)$ is the superconducting transition temperature at $P_c$. Values of $\gamma_I$ are plotted as a function of pressure for measured or estimated $I_c(T,P)$ at 0 K (squares) and 1.0 K (circles). **c,** The ratio between the critical current and SC transition temperature, $I_{c,0}/T_{c,0}$, plotted together with the dc conductivity at $T_c$ onset, $\sigma_{dc}$, as a function of the pressure difference $P-P_c$, where $P_c$ = 2.35 GPa is the QCP. **d,** A contour plot of $I_{c,0}/T_{c,0}$ displayed in the temperature ($T$) and pressure ($P-P_c$) plane. The bold arrow marks $P_c^*$, the boundary between a phase of coexisting antiferromagnetic and superconductivity and a solely superconducting phase, where there is a notable peak and a sudden enhancement in $I_c/T_{c,0}$ for Rh115.

**Supplementary References**

[1] Park, T. & Thompson, J. D. Magnetism and superconductivity in strongly correlated CeRhIn$_5$. *New. J. Phys.* **11**, 055062 (2009).

[2] Dew-Hughes, D. Flux pinning mechanisms in type-II superconductors. *Philos. Mag.* **30**, 293-305 (1974).

[3] Kramer, E. J. Scaling laws for flux pinning in hard superconductors. *J. Appl. Phys.* **44**, 1360-1370 (1973).